\documentclass[12pt,epsfig]{article}
\def \b{{\cal B}}
\def \bea{\begin{eqnarray}}
\def \beq{\begin{equation}}
\def \bo{B^0}

\def \cn{Collaboration}
\def \eea{\end{eqnarray}}
\def \eeq{\end{equation}}
\def \ite{{\it et al.}}

\def \ob{\overline{B}^0}
\def \ok{\overline{K}^0}

\def \s{\sqrt{2}}

\begin{document}

\begin{flushright}
TECHNION-PH-2008-39\\
EFI 08-33 \\
arXiv:0812.4796 \\
December 2008 \\
\end{flushright}

\renewcommand{\thesection}{\Roman{section}}
\renewcommand{\thetable}{\Roman{table}}
\centerline{\bf Doubly CKM-suppressed corrections to CP asymmetries in $B^0 \to J/\psi K^0$}
\medskip
\centerline{Michael Gronau}
\centerline{\it Physics Department, Technion -- Israel Institute of Technology}
\centerline{\it 32000 Haifa, Israel}
\medskip
\centerline{Jonathan L. Rosner}
\centerline{\it Enrico Fermi Institute and Department of Physics}
\centerline{\it University of Chicago, Chicago, Illinois 60637}
\bigskip

\begin{quote}

A doubly CKM-suppressed amplitude in $B^0\to J/\psi K_S$ leads to corrections
in CP asymmetries $S = \sin 2\beta, C=0$, which may be enhanced by
long-distance rescattering.
It has been suggested that this enhancement may lead to several percent corrections.
We calculate an upper bound of order $10^{-3}$ on rescattering corrections using 
measured branching ratios for charmless
$|\Delta S|=1$ $B^0$ decays. 

\end{quote}

\leftline{\qquad PACS codes:  12.15.Hh, 12.15.Ji, 13.25.Hw, 14.40.Nd}

\section{Introduction}
The success of the Kobayashi-Maskawa (KM) model of CP violation
\cite{Kobayashi:1973fv} in predicting correctly CP asymmetries in $B$ meson
decays has been recently recognized by the Nobel committee~\cite{nobel}. The
large asymmetry measured in $B^0\to J/\psi K_{S(L)}$~\cite{Chen:2006nk,:2008cp}
given by $\sin 2\beta$, where $\beta \equiv \phi_1={\rm arg}
(-V^*_{cb}V_{cd}/V^*_{tb}V_{td})$, proved unambiguously that the KM phase is 
the dominant source of CP violation in $B$ decays. 

This test involves an interference between $B^0$-$\bar B^0$ mixing and a $B^0$
decay amplitude~\cite{Carter:1980hr}, consisting of a dominant color-suppressed
$\bar b\to \bar cc\bar s$ tree amplitude and a small contribution from a $\bar
b\to \bar s u \bar u$ penguin amplitude~\cite{Gronau:1989ia,London:1989ph,%
Grinstein:1989df}.  (We use the unitarity of the Cabibbo-Kobayashi-Maskawa
\cite{Kobayashi:1973fv,Cabibbo:1963yz}  (CKM) matrix, $V^*_{tb}V_{ts}=-V^*_{cb}
V_{cs} - V^*_{ub}V_{us}$, 
and will not discuss order $10^{-3}$ effects due to CP violation in 
$B^0-\ob$ mixing~\cite{Bigi:1987in} and $K^0-\ok$ mixing~\cite{Xing:1995jg}.)
The magnitude $\xi$  of the ratio of these two
amplitudes, which determines the theoretical precision of this test, involves 
three suppression factors: A ratio of CKM matrix elements, $|V^*_{ub}V_{us}|/
|V^*_{cb}V_{cs}|\simeq 0.02$~\cite{Amsler:2008zz}, small Wilson coefficients
of penguin operators in the effective Hamiltonian, $c_i\sim 0.04~(i=3,4,5,6)$
(or a QCD loop factor),
and a suppression by the Okubo-Zweig-Iizuka (OZI) rule~\cite{OZI}. The
parameter $\xi$ is expected to be somewhat larger than the product of these 
three factors due to color-suppression of the dominant tree amplitude which
normalizes $\xi$.  Thus, with $\xi \sim 10^{-3}$, it has been commonly accepted
that the measurement of $\sin 2\beta$ in $B^0\to J/\psi K_{S(L)}$ may involve
only a very small uncertainty at a level of $10^{-3}$, or at most a fraction of
a percent~\cite{Gronau:1989ia}. This estimate was supported by calculations of
$\xi$ using QCD factorization~\cite{Boos:2004xp} and perturbative
QCD~\cite{Li:2006vq}.  These perturbative calculations are based on the
absorptive part of the $u$-quark loop, assuming a sufficiently large momentum
transfer in the loop,
and applying rather crude methods for evaluating the four quark operator matrix 
element $\langle J/\psi K^0|(\bar cT^ac)_V(\bar bT^as)_{V-A}|B^0\rangle$.
The absorptive part of the quark loop was proposed thirty years ago as a 
mechanism producing a strong phase leading to CP violation in charged $B$
decays~\cite{Bander:1979px}.
 
The introduction of a small penguin amplitude, carrying a weak phase $\gamma$
and a strong phase $\delta$ relative to the dominant tree amplitude, affects
the time-dependent CP asymmetry in $B^0\to J/\psi K_S$,
\beq
A_{\rm CP}(t)\equiv \frac{\Gamma(\ob \to J/\psi K_S) - \Gamma(\bo\to J/\psi K_S)}
{\Gamma(\ob \to J/\psi K_S) + \Gamma(\bo\to J/\psi K_S)}
= -C\cos(\Delta mt) + S\sin(\Delta mt)~,
\eeq
where, dropping terms quadratic in $\xi$~\cite{Gronau:1989ia},
\beq
C = -2\xi\sin\delta\sin\gamma~~,~~~~~
\Delta S\equiv S-\sin 2\beta = 2\xi\cos 2\beta\cos\delta\sin\gamma~.
\eeq
In the limit $\xi=0$ one has $C=0, \Delta=0$. Current measurements of the two 
asymmetries~\cite{Chen:2006nk,:2008cp,Barberio:2008fa} (based on all charmonium
decays), 
\beq
C(J/\psi K^0)=0.005\pm 0.019~,~~~~~S(J/\psi K^0) = 0.671 \pm 0.024~,
\eeq
involve experimental errors at a level of $\pm 0.02$. This error is
considerably larger than the theoretical uncertainty introduced by the above
estimate of the parameter $\xi$. 

The estimate $\xi \sim 10^{-3}$ has been questioned and challenged in
Refs.~\cite{Ciuchini:2005mg} and \cite{Faller:2008zc}, arguing that 
the $u$-quark penguin amplitude in $B^0\to J/\psi K^0$ may be enhanced by long 
distance rescattering effects from intermediate $S=1$ charmless states to
$J/\psi K^0$.  A sizable enhancement of a penguin amplitude beyond a
perturbative calculation, argued to be due to a large ``charming penguin"
contribution~\cite{Ciuchini:1997hb}, has been observed in $B\to K\pi$. 
It was therefore argued~\cite{Ciuchini:2005mg,Faller:2008zc} that similar 
nonperturbative rescattering effects of intermediate charmless states may
enhance $\xi$, leading to a hadronic uncertainty in $\xi$ at a level of several
percent.

An uncertainty in $\xi$ at this level implies theoretical uncertainties in the
asymmetries $C(J/\psi K^0)$ and $S(J/\psi K^0)$ which are comparable to or even
larger than the current experimental errors in these measurements.  It was
pointed out in Refs.~\cite{Ciuchini:2005mg,Faller:2008zc} that CP asymmetries
$C$ and $\Delta S$ in $B^0\to J/\psi \pi^0$, proportional to $\xi$ in the
flavor SU(3) limit, are enhanced by a factor
$(1-\lambda^2)/\lambda^2=18.6~(\lambda= 0.2257$~\cite{Amsler:2008zz}) relative
to $C$ and $\Delta S$ in $B^0\to J/\psi K_S$.  Thus, it was suggested to study
$\xi$ in $B^0\to J/\psi \pi^0$. Unfortunately, the decay rate for this process
is suppressed by $2\lambda^2/(1-\lambda^2)$ relative to that of $B^0\to J/\psi
K_S$. Consequently one expects errors in the $B^0\to J/\psi\pi^0$ asymmetries
to be correspondingly larger than in the $B^0\to J/\psi K_S$ asymmetries.
Indeed, current measurements~\cite{Barberio:2008fa,:2007wd,Aubert:2008bs}, 
$C(J/\psi\pi^0)= -0.10\pm 0.13, S(J/\psi \pi^0)=-0.93\pm 0.15$, are not
sufficiently accurate for providing useful information about $\xi$. Values of
$\xi$ as large as a few percent cannot be ruled out by the two asymmetries.
 
The purpose of this Letter is to calculate upper bounds on $\xi$ in $B^0\to
J/\psi K^0$ from long distance rescattering effects mediated by charmless
intermediate states.  This provides a re-evaluation of the contribution of the
absorptive part associated with cutting the $u$-quark penguin loop
\cite{Boos:2004xp, Li:2006vq,Bander:1979px} by computing explicitly contributions
of  charmless intermediate states. Using the rich amount of data for numerous 
charmless $B$ meson decays obtained in
experiments at $e^+e^-$ $B$ factories, we will show that the estimate $\xi\sim
10^{-3}$ is much more reasonable than values of $\xi$ at a level of a few
percent.

\section{Upper bounds on rescattering in $B^0\to J/\psi K^0$}
We write the $S$ matrix in terms of $S_0$, which includes strong and
electromagnetic interactions, and $T$, taken to be Hermitian, which corresponds
to the effective weak Hamiltonian at a low energy scale,
\beq
S = S_0 + iT~.
\eeq   
Unitarity of the $S$ matrix $S^\dagger S=1$ implies to first order in $T$,
\beq
T = S_0\,T\,S_0~.
\eeq
Taking matrix elements of the two sides between a $B$ meson state and a final decay state 
$f_0$, and inserting a complete set of intermediate states $f$, one has
\beq\label{unitarity}
\langle f_0|T|B\rangle = \Sigma_f \langle f_0|S_0|f\rangle\langle f|T|B\rangle~,
\eeq
where we used the fact that $B$ is an eigenstate of $S_0$. The matrix elements 
$\langle f_0|T|B\rangle$ and $\langle f|T|B\rangle$ are weak decay amplitudes,
often denoted $A(B\to f_0)$ and $A(B\to f)$, while $\langle f_0|S_0|f\rangle$
represents a rescattering amplitude from $f$ to $f_0$.

Let us first consider the matrix element between $B=B^0$ and $f_0=J/\psi K^0$
for the effective $|\Delta S|=1, \Delta C=0$ operator $T^u$ involving a CKM
factor $V^*_{ub}V_{us}$, 
\beq\label{sum-u}
\langle J/\psi K^0|T^u|B^0\rangle = \Sigma_f \langle J/\psi K^0|S_0|
f \rangle\langle f|T^u|B^0\rangle~.
\eeq
Because $B$ is a spinless particle the $J/\psi$ and $K^0$ are in a $P$-wave.
Consequently the states $f$ are all $J=0, P=-1$ $S=1$  states. Since we are
replacing the absorptive part associated with cutting the $u$-quark penguin
loop by contributions of physical intermediate states, we consider only
charmless states.  This includes a long list of states, such as $f=K^{*+}\pi^-,
\rho^-K^+, K^{*0}\pi^0, \rho^0 K^0$, $\omega K^0, K^{*0}\eta, K^{*0}\eta'$,
and $K^{*+}_0(1430)\pi^-$, but excludes 
$K^+\pi^-, K^0\pi^0$ in an S-wave state and $K^{*+}\rho^-, K^{*0}\rho^0$ in 
$S$ and $D$ waves which have $P=+1$. 

Parity and time-reversal symmetry of $S_0$ imply a reciprocal detailed-balance
relation (we are assuming a single polarization state because $J=0$),
\beq\label{DB}
|\langle J/\psi K^0|S_0|f\rangle| = |\langle f|S_0|J/\psi K^0\rangle|~.
\eeq
Upper bounds on matrix elements $ |\langle f|S_0|J/\psi K^0\rangle|$ for
each of the above states, $f=K^{*+}\pi^-,..., K^{*+}_0(1430)\pi^-$, may be 
obtained using the following considerations. 

We apply Eq.~(\ref{unitarity}) to matrix elements
between $B^0$ and the above charmless final states $f$ for the effective 
$|\Delta S|=1, \Delta C=0$ operator $T^c$ involving a CKM factor $V^*_{cb}V_{cs}$, 
\beq
\langle f|T^c|B^0\rangle = \Sigma_k \langle f|S_0|k \rangle\langle k|T^c|B^0\rangle~.
\eeq
The left-hand-side is dominated by a penguin amplitude which obtains a sizable
``charming penguin" contribution~\cite{Ciuchini:1997hb}. Assuming that a single
intermediate 
state $k=D^{*-}D^+_s$ can at most saturate the sum on the right-hand-side, one has
\beq\label{boundDD}
|\langle f|S_0|D^{*-}D^+_s \rangle||\langle D^{*-}D^+_s|T^c|B^0\rangle| \le
|\langle f|T^c|B^0\rangle|~.
\eeq

On the left-hand-side of (\ref{boundDD}) we can replace the $D^{*-}D^+_s$ state
by $J/\psi K^0$.  One expects $|\langle f|S_0|J/\psi K^0 \rangle| < |\langle
f|S_0|D^{*-}D^+_s \rangle|$ because the first amplitude is OZI-suppressed. 
To calculate the ratio of $B^0$ decay amplitudes into $D^{*-}D^+_s$
and $J/\psi K^0$ we use the expression for decay rates, 
\beq \label{eqn:gamma}
\Gamma = \frac{p_f^*}{8 \pi M_B^2} |\langle f|T|B^0\rangle|^2~,
\eeq
where $p^*_f$ is the momentum of one of the two outgoing particles in the $B^0$
rest frame.  The measured branching ratios and the corresponding momenta
are~\cite{Amsler:2008zz}
\bea
\b(B^0\to D^{*-}D^+_s) &=&  (8.3\pm 1.1)\times 10^{-3}~,
~~~~~~p^*_{D^+_s}=1735~{\rm MeV}/c~,
\nonumber\\
\b(B^0\to J/\psi K^0) &=& (8.71\pm 0.32)\times 10^{-4}~,
~~~p^*_{K^0}=1683~{\rm MeV}/c~.
\eea
This leads to $|\langle D^{*-}D^+_s|T^c|B^0\rangle|/|\langle J/\psi 
K^0|T^c|B^0\rangle| = 3.04 \pm 0.21$. Using the central value, 
Eq.~(\ref{boundDD}) may be replaced by
\beq\label{ineq-13}
|\langle f|S_0|J/\psi K^0 \rangle| < \frac{1}{3}\frac{|\langle f|T^c|B^0\rangle|}
{|\langle J/\psi K^0|T^c|B^0\rangle|}~.
\eeq

We denote
\beq
r_f \equiv \frac{|\langle f|T^u |B^0\rangle|}{|\langle f|T^c |B^0\rangle|}~,
\eeq
and note that $\langle f|T^c|B^0\rangle$ is approximately the total $B^0$ decay
amplitude into $f$,  $\langle f|T|B^0\rangle$; similarly $\langle J/\psi
K^0|T^c|B^0\rangle \approx \langle J/\psi K^0|T|B^0\rangle$.
Combining Eqs.~(\ref{DB}) and (\ref{ineq-13}),  one then obtains the following
upper bound on each of the terms contributing to the sum in (\ref{sum-u}),
normalized by the $B^0$ decay amplitude into $J/\psi K^0$:
\beq\label{bound-xif}
\xi_f \equiv
\frac{| \langle J/\psi K^0|S_0|f \rangle\langle f|T^u|B^0\rangle|}
{|\langle J/\psi K^0|T|B^0\rangle|} < \frac{1}{3}r_f
\left(\frac{|\langle f|T|B^0\rangle|}
{|\langle J/\psi K^0|T|B^0\rangle|}\right )^2~.
\eeq  
This upper bound is a central result in our analysis.
It should be considered a strong inequality (in which the factor $1/3$ may 
be replaced by $1/10$) because it is based on a conservative 
inequality (\ref{boundDD}) and on presumably strong OZI-suppression of 
$|\langle f|S_0|J/\psi K^0 \rangle|$ relative to $|\langle f|S_0|D^{*-}D^+_s \rangle|$.

\section{Numerical upper bounds on rescattering}
We now study numerical bounds on rescattering parameters $\xi_f$ for numerous 
intermediate states $f$ in $B^0\to f \to J/\psi K^0$. We start by discussing $S=1$ 
charmless states $f=VP$ consisting of pairs of vector and pseudoscalar mesons. 
We use
\beq\label{p3}
\left(\frac{|\langle f|T|B^0\rangle|}
{|\langle J/\psi K^0|T|B^0\rangle|}\right )^2 = 
\frac{\b(B^0\to f)}{\b(B^0\to J/\psi K^0)}\left(\frac{p^*_{K^0}}{p^*_f}\right)~.
\eeq

Values for the parameter $r_f$, the ratio of two amplitudes in $B\to VP$
involving CKM factors $V^*_{ub}V_{us}$ and $V^*_{cb}V_{cs}$, are extracted from
a study applying broken flavor SU(3) to these decays and decays into corresponding
$S=0$ charmless states~\cite{Chiang:2003pm}. In the language of
Ref.~\cite{Dighe:1997wj}, matrix elements $\langle f|T^u |B^0\rangle$
involve combinations of graphical amplitudes representing color-favored and
color-suppressed tree amplitudes $T'_{V(P)}$ and $C'_{V(P)}$, while $\langle
f|T^c |B^0\rangle$ involve penguin amplitudes $P'_{V(P)}$, singlet penguin
amplitudes $S'_{V(P)}$ (corresponding to SU(3) singlet mesons in the final
state), and electroweak penguin amplitudes $P'_{EW,V(P)}$.  
The subscript $V$ or $P$ denotes the final-state meson (vector or pseudoscalar)
incorporating the spectator quark.  We are neglecting color-suppressed
electroweak penguin contributions.  SU(3) breaking is included in $T'_V$ and
$T'_P$ in terms of ratios of pseudoscalar and vector meson decay constants,
$f_K/f_{\pi}$ and $f_{K^*}/f_{\rho}$, respectively.  Expressions for the above
matrix elements are given in Table I for $B^0$ decays into seven $VP$ states.
Note that that while $\langle f|T^u |B^0\rangle$ involves
color-allowed tree amplitudes $T'_P$ and $T'_V$ for $f=K^{*+}\pi^-$ and
$f=\rho^-K^+$, it is governed by color-suppressed amplitudes $C'_P$ and $C'_V$
for all other final states. Consequently the  values of $r_f$ in the first two
processes are expected to be considerably larger than in the others. 
%
\begin{table}
\caption{Expressions for matrix elements $\langle f|T^u |B^0\rangle$
and $ \langle f|T^c |B^0\rangle$ in $B^{0}\to VP$ in terms of graphical amplitudes.
\label{tab:amps}}
\begin{center} 
\begin{tabular}{c c c} \hline \hline
Final state & $\langle f|T^u|B^0\rangle$  & $ \langle f|T^c |B^0\rangle$ \\
\hline
$K^{*+}\pi^-$ & $-T'_P$  & $ -P'_P$  \\
$\rho^- K^+$    & $-T'_V$  &   $-P'_P$  \\
$K^{*0}\pi^0$ & $-C'_V/\s$  & $(P'_P - P'_{EW,V})/\s$ \\
$\rho^0 K^0$ &  $-C'_P/\s$  &  $(P'_V - P'_{EW,P})/\s$ \\
$\omega K^0$  & $C'_P/\s$  & $(P'_V + 2S'_P + \frac{1}{3}P'_{EW,P})/\s$ \\
$K^{*0}\eta$  & $-C'_V/\sqrt{3}$  &   $(P'_V - P'_P - S'_V - \frac{2}{3}P'_{EW,V})/\sqrt{3}$ \\
$K^{*0}\eta'$ & $C'_V/\sqrt{6}$  &  $ (2P'_V + P'_P + 4S'_V - \frac{1}{3}P'_{EW,V})/\sqrt{6}$ \\
\hline \hline
\end{tabular}
\end{center}
\end{table}

We calculate numerical values for $r_f$ using entries in the third column of 
Table V in Ref.~\cite{Chiang:2003pm}, updating some values by fitting to more 
recent measurements of $\b(B^0\to K^{*0}\pi^0), \b(B^0\to K^{*0}\eta)$ and 
$\b(B^0\to K^{*0}\eta')$.
Branching ratios for the seven $B^0\to VP$ decays~\cite{Barberio:2008fa}, 
corresponding center-of-mass momenta $p^*$~\cite{Amsler:2008zz}, and values 
of $r_f$ are used to calculate from Eqs.~(\ref{bound-xif}) and (\ref{p3})
upper bounds on $\xi_f$ for the these intermediate $VP$ states. Input values
and resulting upper bounds on $\xi_f$ are summarized in Table \ref{tab:xi-f}.
The largest upper bounds, $\xi_f <(7.9\pm 1.1)\times 10^{-4}$ and 
$\xi_f<(5.6\pm 1.0)\times 10^{-4}$, are obtained for $f=K^{*+}\pi^-$ and $f=\rho^-K^+$,
respectively. Much smaller values, at a level of $10^{-4}$, are calculated for all other 
$VP$ states. 

\begin{table}
\caption{Branching ratios~\cite{Barberio:2008fa}, center-of-mass
momenta~\cite{Amsler:2008zz}, parameters $r_f$ and upper bounds on $\xi_f$ for
seven charmless intermediate $VP$ states.
\label{tab:xi-f}}
\begin{center}
\begin{tabular}{c c c c c} \hline \hline
    Mode    &   $\b~$   & $p^*$ &    
$r_f$      &     Upper bound on $\xi_f$   \\
    $f$       &  $(10^{-6})$ & (MeV) &     &   $(10^{-4}$)\\
\hline
$K^{*+}\pi^-$ & 10.3$\pm$1.1 & 2563 & 0.31$\pm$0.03 & 7.9$\pm$1.1 \\
$\rho^- K^+$    &  8.6$\pm$1.0 & 2559 & 0.26$\pm$0.03 & 5.6$\pm$1.0 \\
$K^{*0}\pi^0$ &  2.4$\pm$0.7 & 2562 & 0.09$\pm$0.04 & 0.6$\pm$0.3 \\
$\rho^0 K^0$    &  5.4$\pm$1.0 & 2558 & 0.04$\pm$0.03 & 0.5$\pm$0.4 \\
$\omega K^0$  &  5.0$\pm$0.6 & 2557 & 0.04$\pm$0.03 & 0.5$\pm$0.4 \\
$K^{*0}\eta$  & 15.9$\pm$1.0 & 2534 & 0.04$\pm$0.02 & 1.6$\pm$0.7 \\
$K^{*0}\eta'$ &  3.8$\pm$1.2 & 2471 & 0.08$\pm$0.04 & 0.8$\pm$0.4 \\
\hline \hline
\end{tabular}
\end{center}
\end{table}

Because rescattering effects of the form $B^0\to f \to J/\pi K^0$ increase with 
$\b(B^0 \to f)$, we search for charmless intermediate states $f$ for which this
branching ratio is particularly large. We note that the three-body decay mode
$B^0\to K^0\pi^+\pi^-$, with $\b=(44.8\pm 2.6)\times 10^{-6}$ is dominated by
the quasi-two-body decay $B^0\to K^*_0(1430)^+\pi^-$, involving a scalar and a
pseudoscalar meson in a $P=-1$ $S$-wave~\cite{Amsler:2008zz}: 
\beq\label{BR-K*0}
\b[B^0\to K^*_0(1430)^+\pi^-] = (50^{+8}_{-9})\times 10^{-6}~.
\eeq
The fact that this branching ratio seems to exceed that for the three-body
final state indicates strong destructive interference with other amplitudes
including $B^0\to K^{*+}\pi^-,~\rho^0K^0,~f_0(980)K^0$, and a 
non-resonant amplitude~\cite{Garmash:2006fh}.

In order to evaluate an upper bound for $\xi_f$ based on the $u$-quark
amplitude's contribution to the $K^*_0(1430)^+\pi^-$ intermediate state, we
must obtain an estimate of the value of $r_f$ for this state.  This quantity
(the subscript $P$ denotes the final pseudoscalar meson incorporating the
spectator quark),
\beq\label{rfK*0}
r_f \equiv \frac{|T'_P(B^0\to K^{*+}_0\pi^-)|}{|P'_P(B^0\to K^{*+}_0\pi^-)|}~,
\eeq
 is the ratio of the $u$-quark tree amplitude and the $c$-quark penguin 
amplitude in $B^0 \to K^*_0(1430)^+\pi^-$. While the latter amplitude 
dominates this process, the former may be estimated to a good approximation 
assuming factorization. A similar situation occurs in $B^0\to K^+\pi^-$, where
the ratio of tree and penguin amplitudes has been determined within a global
flavor SU(3) fit to all $B\to K\pi$ and $B\to\pi\pi$
decays~\cite{Chiang:2004nm},
\beq\label{T/P}
\frac{|T'(B^0\to K^+\pi^-)|}{|P'(B^0\to K^+\pi^-)|} = 
\frac{0.281(16.1\pm 2.0)}{48.2 \pm 1.0} = 0.094 \pm 0.012~.
\eeq

The ratio of the two penguin amplitudes dominating $B^0\to K^+\pi^-$ and 
$B^0\to K^*_0(1430)^+\pi^-$ is obtained from the corresponding partial rates,
\beq\label{P/PP}
\frac{|P'(B^0\to K^+\pi^-)|}{|P'_P(B^0\to K^{*+}_0\pi^-)|} \approx 
\sqrt{\frac{\Gamma(B^0\to K^+\pi^-)}{\Gamma(B^0\to K^{*+}_0\pi^-)}
\frac{p^*(K^{*+}_0)}{p^*(K^+)}} = 0.60 \pm 0.05~.
\eeq  
Here we have used (\ref{BR-K*0}) with~\cite{Amsler:2008zz} $\b(B^0\to K^+\pi^-)
 = (19.4 \pm 0.6)\times 10^{-6},~p^*(K^+, K^{*+}_0)=(2615,2445)$ MeV$/c$.

In the factorization approximation the tree amplitudes $T'(B^0\to K^+\pi^-)$
and $T'_P(B^0\to K^{*+}\pi^-)$ involve respectively the $K$ and $K^*_0$ decay
constants, $f_K$ and $f_{K^*_0}$, and the $B$ to $\pi$ form factors at
$q^2=m^2_K$ and $m^2_{K^*_0}$ which are assumed to be approximately equal.
Thus,
\beq\label{TP/T}
\frac{|T'_P(B^0\to K^{*+}\pi^-)|}{|T'(B^0\to K^+\pi^-)|}
\approx \frac{f_{K^*_0}}{f_K}~.
\eeq
Note that the scalar $K^*_0$ couples to the weak vector current through a
coupling proportional to the $K^*_0$ decay constant $f_{K^*_0}$ which vanishes
by $G$-parity in the SU(3) symmetry limit~\cite{Laplace:2001qe,Diehl:2001xe,%
Cheng:2005nb}.  SU(3) breaking leads to a nonzero value, expected to be of
order $(m_s-m_d)/\Lambda_{\rm QCD}$ relative to the $K$ meson decay constant.
Theoretical calculations of $f_{K^*_0}$ lead to values in the
range~\cite{Maltman:1999jn,Narison:1999uj,Shakin:2001sz,Du:2004ki} 
\beq\label{fK*0}
f_{K^*_0} = 40 \pm 6~{\rm MeV}~,
\eeq
to be compared with $f_K=155.5 \pm 0.8$ MeV~\cite{JLRST}.

Taking a product of the three factors in (\ref{T/P}), (\ref{P/PP}), and
(\ref{TP/T}), we find
\beq
r_f = 0.015 \pm 0.003~,~~~~~f=K^*_0(1430)^+\pi^-~.
\eeq
An upper bound on $\xi_f$ for $f=K^*_0(1430)^+\pi^-$ is then obtained using 
Eqs.~(\ref{bound-xif}) and (\ref{p3}) while taking into account correlated
errors,
\bea\label{xfK*0}
\xi_f  & < & \frac{1}{3}r_f \frac{\b[B^0 \to K^*_0(1430)^+ \pi^-]}
{\b(B^0 \to J/\psi K^0)} \frac{p^*(K^0)}{p^*(K^{*+}_0)} 
\nonumber \\
& = & (1.9 \pm 0.4) \times 10^{-4}~,~~~~~~f=K^*_0(1430)^+ \pi^-~.
\eea
Thus, in spite of the large branching ratio measured for this decay mode, this
upper bound is about four times smaller than the largest value obtained for 
the corresponding $VP$ state $K^{*+}\pi^-$ in Table \ref{tab:xi-f}.  

There are good prospects for replacing the theoretical estimate (\ref{fK*0})
with an experimentally determined value.  The partial width for the decay
$\tau^- \to M^- \nu$, where $M$ is a strange scalar or pseudoscalar meson, is
\beq \label{eqn:td}
\Gamma(\tau^- \to M^- \nu) = G_F^2 |V_{us}|^2 \frac{f_M^2}{16 \pi}
\frac{(m_\tau^2 - m_M^2)^2}{m_\tau}~.
\eeq
With $G_F = 1.16637(1) \times 10^{-5}$ GeV$^{-2}$, $m_\tau = 1.77684(17)$
GeV/$c^2$, $m_K = 0.493677(16)$ GeV/$c^2$, $|V_{us}| = 0.2255(19)$, and
$f_K = 0.1555(8)$ GeV \cite{Amsler:2008zz} (for the last two, see Ref.\
\cite{JLRST}), this yields a prediction $\Gamma(\tau^- \to K^- \nu) = (1.59 \pm
0.03) \times 10^{-14}$ GeV.  The lifetime of the $\tau$ is $(290.6 \pm 1.0)
\times 10^{-15}$ s \cite{Amsler:2008zz}, implying the prediction $\b(\tau^- \to
K^- \nu) = (7.02 \pm 0.14) \times 10^{-3}$, in satisfactory agreement with the
experimental value \cite{Amsler:2008zz} $(6.95 \pm 0.23) \times 10^{-3}$.

The prediction (\ref{eqn:td}) can be applied to the $K^*_0(1430)$ in the
narrow-width approximation, permitting one to obtain the ratio of scalar and
pseudoscalar decay constants:
\beq
\frac{f_{K^*_0}}{f_K} = \frac{m_\tau^2 - m_K^2}{m_\tau^2 - m_{K^*_0}^2}
\sqrt{\frac{\b(\tau \to K^*_0 \nu)}{\b(\tau \to K \nu)}}~.
\eeq
With the experimental upper limit \cite{Barate:1999hj} $\b[\tau \to K^*_0(1430)
\nu] < 5 \times 10^{-4}$ (95\% c.l.) and using the central value of
$m[K^*_0(1430)] = 1425 \pm 50$ MeV/$c^2$ \cite{Amsler:2008zz}, this ratio is
less than 0.69, entailing $f_{K^*_0} < 107.5$ MeV if the predicted value
of $\b(\tau^- \to K^- \nu)$ is used.  The large width $\Gamma[K^*_0(1430)] =
270 \pm 80$ MeV leads to a small positive correction of 1.127 to the predicted
partial width for $\tau \to K^*_0(1430) \nu$, reducing this upper bound
slightly to $f_{K^*_0} < 101$ MeV.  This is about a factor of 2.5 larger than
the theoretical estimates summarized in Eq.\ (\ref{fK*0}).  Using those
estimates and including the finite-width correction, we predict
$\b[\tau \to K^*_0(1430) \nu] = (7.8 \pm 2.3) \times 10^{-5}$.  As stressed
in Ref.\ \cite{Diehl:2001xe}, this should be accessible in present
experiments.

The intermediate state $f=K^*_0(1430)^0\pi^0$ is fed by a color-suppressed 
$u$-quark tree amplitude. Its penguin-dominated branching ratio is expected to
be about half of that measured for $B^0\to K^*_0(1430)^+\pi^-$. Consequently
the upper bound on $\xi_f$ for $f=K^*_0(1430)^0\pi^0$ is considerably smaller
than (\ref{xfK*0}).

\section{Conclusion}
We have calculated upper bounds on contributions to the doubly-CKM-suppressed
parameter $\xi$ from rescattering $B^0\to f \to J/\psi K^0$ through charmless
$S=1$ intermediate states $f$.  We have derived in Eq.~(\ref{bound-xif}) a
general conservative upper bound on $\xi_f$ which increases with $\b(B^0\to f)$ 
and with the ratio $r_f$ of tree and penguin amplitudes in $B^0\to f$.  
The actual upper bound may involve a factor $1/10$ instead of $1/3$ because OZI
suppression in $f \to J/\psi K^0$ has not been included in (\ref{bound-xif}).

The highest upper bounds on $\xi_f$,  somewhat below $10^{-3}$, were obtained
for $f=K^{*+}\pi^-$ and $\rho^- K^+$, while other intermediate states with
neutral vector and pseudoscalar mesons involve much smaller rescattering
contributions. This applies also to the state $K^*_0(1430)^+\pi^-$, which has
the largest quasi-two-body decay branching ratio measured so far in $B^0$
decays, $\b[B^0\to K^*_0(1430)^+\pi^-] = (50^{+8}_{-9})\times 10^{-6}$.
We noted destructive interference between $B^0\to K^*_0(1430)^+\pi^-$ and 
other modes contributing to $B^0\to K^0\pi^+\pi^-$.  This indicates potential
destructive interference between rescattering contributions to $\xi$ of these
intermediate states.

One may wonder whether larger contributions to $\xi$ may originate in charmless
intermediate states with multiplicity larger than three.
Only two $S=1$ charmless branching ratios comparable to that of 
$B^0\to K^*_0(1430)^+\pi^-$ have been measured~\cite{Aubert:2007fm},
$\b(B^0\to K^{*0}\pi^+\pi^-)=(54.5 \pm 5.2)\times 10^{-6},
\b(B^0\to K^{*0}K^+K^-)=(27.5\pm 2.6)\times 10^{-6}$.
These involve quasi-three-body decays leading to four particles in the final
state.  $P=-1$ projections of these states, with smaller branching ratios, may
rescatter into $J/\psi K^0$.  Although it is difficult to calculate or measure
the tree-to-penguin ratio $r_f$ for these states, we expect it to be no more
than $0.1$. The isospin relation~\cite{Gronau:2005ax} $\Gamma(B^0\to K^{*0}
\pi^+\pi^-)=\Gamma(B^+\to K^{*+}\pi^+\pi^-)$ which holds within
$1.5\sigma$~\cite{Barberio:2008fa} is consistent with $r_f=0$ in these
processes. We do not anticipate constructive interference between rescattering 
contributions of the intermediate states $K^{*0}\pi^+\pi^-$ and $K^{*0}K^+K^-$
and the above calculated contributions of $K^{*+}\pi^-$ and $\rho^-K^+$
which are probably larger.

Rescattering from intermediate states with two vector mesons in a $P=-1$
$P$-wave state, including $K^{*+}\rho^-, K^{*0}\rho^0$ and $K^{*0}\phi$,
involve branching ratios considerably smaller than those of $B^0\to K^{*0}
\pi^+\pi^-$ and $B^0\to K^{*0} K^+K^-$~\cite{Amsler:2008zz} and very small
$u$-quark tree amplitudes.  For instance, the tree amplitude in $B^0\to
K^{*+}\rho^-$ is related by flavor SU(3) to the amplitude dominating $B^0 \to
\rho^+\rho^-$. Approximately $100\%$ longitudinal polarization has been
measured in this process~\cite{Barberio:2008fa}, corresponding to a combination
of $S$ and $D$ waves but no $P$ wave. This implies a negligible $u$-quark $P$
wave amplitude in $B^0\to K^{*+}\rho^-$.  Similarly, the tree amplitude in
$B^0\to K^{*0}\rho^0$ is color-suppressed, while the one in $B^0 \to
K^{*0}\phi$ is both color and OZI-suppressed.  The contributions of two vector
meson intermediate states to $\xi$ are therefore negligible.
 
Thus, we expect a value of $\xi$ which is at most a few times $10^{-3}$, in
agreement with an early estimate~\cite{Gronau:1989ia} and in contrast to a
suggestion for an order of magnitude larger enhancement of $\xi$ by long 
distance effects~\cite{Ciuchini:2005mg,Faller:2008zc}.

\vskip -5mm
\section*{Acknowledgments}
We thank Robert Fleischer, Thomas Mannel, Blazenka Melic and Misha 
Vysotsky for useful discussions.  The work of
J. L. R. was supported in part by the United States Department of Energy
through Grant No.\ DE FG02 90ER40560.

\vskip -5mm
\def \ajp#1#2#3{Am.\ J. Phys.\ {\bf#1}, #2 (#3)}
\def \apny#1#2#3{Ann.\ Phys.\ (N.Y.) {\bf#1}, #2 (#3)}
\def \app#1#2#3{Acta Phys.\ Polonica {\bf#1}, #2 (#3)}
\def \arnps#1#2#3{Ann.\ Rev.\ Nucl.\ Part.\ Sci.\ {\bf#1}, #2 (#3)}
\def \art{and references therein}
\def \cmts#1#2#3{Comments on Nucl.\ Part.\ Phys.\ {\bf#1}, #2 (#3)}
\def \cn{Collaboration}
\def \cp89{{\it CP Violation,} edited by C. Jarlskog (World Scientific,
Singapore, 1989)}
\def \econf#1#2#3{Electronic Conference Proceedings {\bf#1}, #2 (#3)}
\def \efi{Enrico Fermi Institute Report No.}
\def \epjc#1#2#3{Eur.\ Phys.\ J.\ C {\bf#1}, #2 (#3)}
\def \f79{{\it Proceedings of the 1979 International Symposium on Lepton and
Photon Interactions at High Energies,} Fermilab, August 23-29, 1979, ed. by
T. B. W. Kirk and H. D. I. Abarbanel (Fermi National Accelerator Laboratory,
Batavia, IL, 1979}
\def \hb87{{\it Proceeding of the 1987 International Symposium on Lepton and
Photon Interactions at High Energies,} Hamburg, 1987, ed. by W. Bartel
and R. R\"uckl (Nucl.\ Phys.\ B, Proc.\ Suppl., vol. 3) (North-Holland,
Amsterdam, 1988)}
\def \ib{{\it ibid.}~}
\def \ibj#1#2#3{~{\bf#1}, #2 (#3)}
\def \ichep72{{\it Proceedings of the XVI International Conference on High
Energy Physics}, Chicago and Batavia, Illinois, Sept. 6 -- 13, 1972,
edited by J. D. Jackson, A. Roberts, and R. Donaldson (Fermilab, Batavia,
IL, 1972)}
\def \ijmpa#1#2#3{Int.\ J.\ Mod.\ Phys.\ A {\bf#1}, #2 (#3)}
\def \ite{{\it et al.}}
\def \jhep#1#2#3{JHEP {\bf#1} (#3) #2}
\def \jpb#1#2#3{J.\ Phys.\ B {\bf#1}, #2 (#3)}
\def \lg{{\it Proceedings of the XIXth International Symposium on
Lepton and Photon Interactions,} Stanford, California, August 9--14, 1999,
edited by J. Jaros and M. Peskin (World Scientific, Singapore, 2000)}
\def \lkl87{{\it Selected Topics in Electroweak Interactions} (Proceedings of
the Second Lake Louise Institute on New Frontiers in Particle Physics, 15 --
21 February, 1987), edited by J. M. Cameron \ite~(World Scientific, Singapore,
1987)}
\def \kaon{{\it Kaon Physics}, edited by J. L. Rosner and B. Winstein,
University of Chicago Press, 2001}
\def \kdvs#1#2#3{{Kong.\ Danske Vid.\ Selsk., Matt-fys.\ Medd.} {\bf #1}, No.\
#2 (#3)}
\def \ky{{\it Proceedings of the International Symposium on Lepton and
Photon Interactions at High Energy,} Kyoto, Aug.~19-24, 1985, edited by M.
Konuma and K. Takahashi (Kyoto Univ., Kyoto, 1985)}
\def \mpla#1#2#3{Mod.\ Phys.\ Lett.\ A {\bf#1}, #2 (#3)}
\def \nat#1#2#3{Nature {\bf#1}, #2 (#3)}
\def \nc#1#2#3{Nuovo Cim.\ {\bf#1}, #2 (#3)}
\def \nima#1#2#3{Nucl.\ Instr.\ Meth.\ A {\bf#1}, #2 (#3)}
\def \np#1#2#3{Nucl.\ Phys.\ {\bf#1}, #2 (#3)}
\def \npps#1#2#3{Nucl.\ Phys.\ Proc.\ Suppl.\ {\bf#1}, #2 (#3)}
\def \os{XXX International Conference on High Energy Physics, Osaka, Japan,
July 27 -- August 2, 2000}
\def \PDG{Particle Data Group, D. E. Groom \ite, \epjc{15}{1}{2000}}
\def \pisma#1#2#3#4{Pis'ma Zh.\ Eksp.\ Teor.\ Fiz.\ {\bf#1}, #2 (#3) [JETP
Lett.\ {\bf#1}, #4 (#3)]}
\def \pl#1#2#3{Phys.\ Lett.\ {\bf#1}, #2 (#3)}
\def \pla#1#2#3{Phys.\ Lett.\ A {\bf#1}, #2 (#3)}
\def \plb#1#2#3{Phys.\ Lett.\ B {\bf#1} (#3) #2}
\def \pr#1#2#3{Phys.\ Rev.\ {\bf#1}, #2 (#3)}
\def \prc#1#2#3{Phys.\ Rev.\ C {\bf#1}, #2 (#3)}
\def \prd#1#2#3{Phys.\ Rev.\ D {\bf#1} (#3) #2}
\def \prl#1#2#3{Phys.\ Rev.\ Lett.\ {\bf#1} (#3) #2}
\def \prp#1#2#3{Phys.\ Rep.\ {\bf#1}, #2 (#3)}
\def \ptp#1#2#3{Prog.\ Theor.\ Phys.\ {\bf#1} (#3) #2}
\def \rmp#1#2#3{Rev.\ Mod.\ Phys.\ {\bf#1}, #2 (#3)}
\def \rp#1{~~~~~\ldots\ldots{\rm rp~}{#1}~~~~~}
\def \si90{25th International Conference on High Energy Physics, Singapore,
Aug. 2-8, 1990}
\def \slc87{{\it Proceedings of the Salt Lake City Meeting} (Division of
Particles and Fields, American Physical Society, Salt Lake City, Utah, 1987),
ed. by C. DeTar and J. S. Ball (World Scientific, Singapore, 1987)}
\def \slac89{{\it Proceedings of the XIVth International Symposium on
Lepton and Photon Interactions,} Stanford, California, 1989, edited by M.
Riordan (World Scientific, Singapore, 1990)}
\def \smass82{{\it Proceedings of the 1982 DPF Summer Study on Elementary
Particle Physics and Future Facilities}, Snowmass, Colorado, edited by R.
Donaldson, R. Gustafson, and F. Paige (World Scientific, Singapore, 1982)}
\def \smass90{{\it Research Directions for the Decade} (Proceedings of the
1990 Summer Study on High Energy Physics, June 25--July 13, Snowmass,
Colorado),
edited by E. L. Berger (World Scientific, Singapore, 1992)}
\def \tasi{{\it Testing the Standard Model} (Proceedings of the 1990
Theoretical Advanced Study Institute in Elementary Particle Physics, Boulder,
Colorado, 3--27 June, 1990), edited by M. Cveti\v{c} and P. Langacker
(World Scientific, Singapore, 1991)}
\def \TASI{{\it TASI-2000:  Flavor Physics for the Millennium}, edited by J. L.
Rosner (World Scientific, 2001)}
\def \yaf#1#2#3#4{Yad.\ Fiz.\ {\bf#1}, #2 (#3) [Sov.\ J.\ Nucl.\ Phys.\
{\bf #1}, #4 (#3)]}
\def \zhetf#1#2#3#4#5#6{Zh.\ Eksp.\ Teor.\ Fiz.\ {\bf #1}, #2 (#3) [Sov.\
Phys.\ - JETP {\bf #4}, #5 (#6)]}
\def \zpc#1#2#3{Zeit.\ Phys.\ C {\bf#1}, #2 (#3)}
\def \zpd#1#2#3{Zeit.\ Phys.\ D {\bf#1}, #2 (#3)}

\end{document}